\documentclass[a4paper,twocolumn,american]{revtex4}
\usepackage{times}
\usepackage[T1]{fontenc}
\usepackage[latin1]{inputenc}
\usepackage{graphicx}

\makeatletter

\providecommand{\tabularnewline}{\\}

\usepackage{babel}
\makeatother
\begin{document}

\title{Remote preparation of an atomic quantum memory}

\author{Wenjamin Rosenfeld$^{1}$, Stefan Berner$^{1}$, Jürgen Volz$^{1}$,
Markus Weber$^{1}$, and Harald Weinfurter$^{1,2}$}

\affiliation{$^{1}$Department für Physik, Ludwig-Maximilians Universität München,
D-80799 München, Germany\\
$^{2}$Max-Planck Institut für Quantenoptik, D-85748 Garching, Germany}

\begin{abstract}
Storage and distribution of quantum information are key elements of
quantum information processing and quantum communication. Here, using
atom-photon entanglement as the main physical resource, we experimentally
demonstrate the preparation of a distant atomic quantum memory. Applying
a quantum teleportation protocol on a locally prepared state of a
photonic qubit, we realized this so-called remote state preparation
on a single, optically trapped $^{87}$Rb
 atom. We evaluated the performance of this scheme by the full tomography
of the prepared atomic state, reaching an average fidelity of 82\%.

\end{abstract}
\maketitle
Quantum teleportation\cite{Teleportations93-04} and quantum cryptography\cite{Gisin02}
were the first quantum communication methods experimentally demonstrated.
Meanwhile, first devices for secure communication became already commercially
available. For the next step of quantum information processing, new
methods and technologies are required. Many new concepts of quantum
information science, for example the quantum repeater\cite{Briegel98}
or quantum networks, all the way towards distributed quantum computing,
require a device interfacing photonic quantum channels and matter-based
quantum memories and processors.

\noindent So far, there are two methods experimentally investigated.
The first employs atomic ensembles to momentarily store quantum states
of light. Recently, qubits encoded on single photons or qunits encoded
in the quantum state of an electromagnetic field have been transferred
to the collective state of atoms and vice versa\cite{Matsukevich04,Polzik04}.
An impressive experimental demonstration of a first quantum communication
protocol, the quantum teleportation of coherent states of light, was
reported very recently\cite{Polzik06}.

\noindent In the second method the desired interface to a photonic
communication channel can be realized using the recently achieved
entanglement between a single atom and a single photon\cite{Blinov04,Volz06}.
This method applies directly to well-studied single quantum systems
like trapped neutral atoms or ions. For linear ion chains and neutral
atoms in optical lattices, various methods of quantum information
storage and processing were already demonstrated, e.g. entanglement
of up to 8 ions\cite{Blatt05,Wineland05}, creation of a cluster state
involving tens of neutral atoms\cite{Mandel03} or manipulations on
a neutral atom quantum shift register\cite{Meschede06}. Furthermore,
this interface concept can be adopted to other qubit systems, like
optically addressed quantum dots\cite{Sham03,Imamoglu05,Lukin06}
or superconducting QED-systems\cite{Schoelkopf04}, stimulating novel
applications in these areas as well. 

Here we report the first experimental realization of a quantum communication
protocol based on atom-photon entanglement. We perform full remote
preparation of an atomic quantum memory via teleportation of an arbitrarily
prepared quantum state of a single photon, using matter-light entanglement
as the interface between the memory device and the communication channel.
This method uses expansion of the Hilbert space of one particle of
the entangled pair with subsequent complete Bell-state analysis. Being
formally equivalent to quantum teleportation\cite{Popescu95,DeMartini98}
it enables the transfer of a known quantum state from the photon to
the atom. Recently, various approaches towards remote state preparation
were studied experimentally with entangled photons\cite{Kwiat05},
light beams\cite{Lvovsky04} and nuclear magnetic spins\cite{Gao03},
however without expansion of the Hilbert space and without complete
Bell-state analysis and thus with significantly reduced performance.

Our experiment includes four steps: (i) Entanglement is generated
between the spin of a single trapped $^{87}$Rb
 atom and the polarization of a single spontaneously emitted photon\cite{Volz06}.
(ii) An additional degree of freedom of the photon is used to encode
the quantum state we wish to transfer\cite{Popescu95}. (iii) The
photon is subject to a complete Bell-state measurement\cite{DeMartini98,Michler00},
projecting the atom into one of four well-defined states. (iv) The
success of the transfer is shown with full quantum state tomography
of the atomic qubit.

\begin{figure}[bt]
\begin{center}\includegraphics{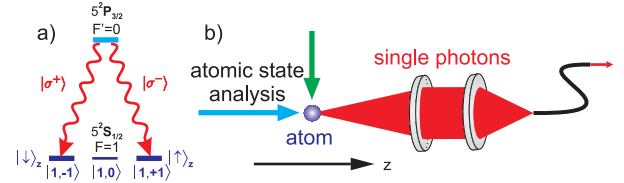}\end{center}

\caption{\label{cap:atomPart}Schematic of atom-photon entanglement generation
in a spontaneous decay of a single optically trapped $^{87}\textrm{Rb}$
atom. (a) After optical excitation to $\mathrm{F'=0}$, the atom decays
into the ground state manifold $\left|\uparrow\right\rangle _{z},\:\left|\downarrow\right\rangle _{z}$
forming an entangled state between the atomic spin and the polarization
of the emitted photon. (b) The emitted photon is collected with a
microscope objective, coupled into a 5 m long single-mode optical
fiber and guided to the preparation setup shown in Fig. \ref{cap:preparationPart}.
The overall detection efficiency for the photon is about $3\cdot10^{-4}$.}
\end{figure}

In more detail, we first establish entanglement between a photon and
a single neutral $^{87}\textrm{Rb}$ atom stored in an optical dipole
trap\cite{Weber06}. Therefore the atom is optically excited to the
$5{}^{2}P_{3/2}$, $\left|F'=0,\, m_{F'}=0\right\rangle $ state (see
Fig. \ref{cap:atomPart} (a)). In the following spontaneous decay
the polarization of the emitted photon is entangled with the spin
state of the atom\cite{Volz06}, resulting in the maximally entangled
state

\begin{equation}
\left|\Psi^{+}\right\rangle =\frac{1}{\sqrt{2}}(\left|\downarrow\right\rangle _{z}\left|\sigma^{+}\right\rangle +\left|\uparrow\right\rangle _{z}\left|\sigma^{-}\right\rangle ),\end{equation}
where $\left|\sigma^{\pm}\right\rangle $ are the right- and left-circular
polarization states of the emitted photon. The two states $\left|\uparrow\right\rangle _{z}$
and $\left|\downarrow\right\rangle _{z}$, defining the atomic qubit,
correspond to the $\left|F=1,\, m_{F}=\pm1\right\rangle $ Zeeman
sublevels of the $5{}^{2}S_{1/2},\: F=1$ hyperfine ground level.

\begin{figure}[bt]
\begin{center}\includegraphics{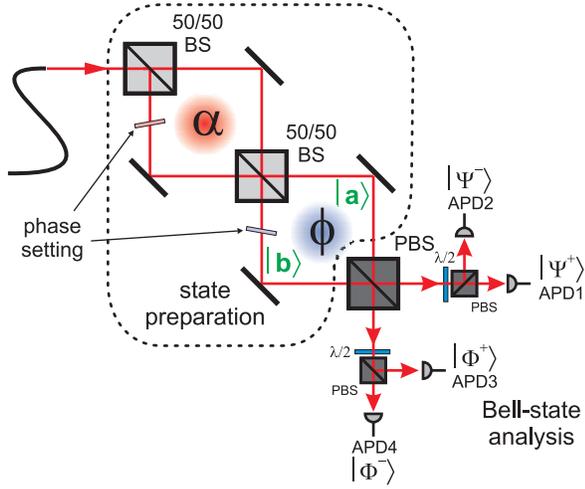}\end{center}

\caption{\label{cap:preparationPart}Schematic setup for preparing the state
from Eq. (\ref{eq:spatialState}) on the spatial degree of freedom
of the photon and for the subsequent Bell-state measurement. The interferometric
phase setting $(\alpha,\phi)$ allows to prepare any desired superposition
of the spatial modes $\left|a\right\rangle $ and $\left|b\right\rangle $
without affecting the polarization degree of freedom. The following
polarizing beam-splitter (PBS) together with the polarization measurement
in $\left|\pm45°\right\rangle $ basis enable a complete Bell-state
analysis in the combined polarization/spatial-mode Hilbert space of
the photon.}
\end{figure}

For the next step the emitted photon is coupled into a single-mode
optical fiber (Fig. \ref{cap:atomPart} (b)) and guided to the setup
shown in Fig. \ref{cap:preparationPart}, where the state we wish
to transfer is imprinted onto the photon. For this purpose we extend
the Hilbert space of the photon by using two spatial modes as an additional
degree of freedom. The photon is coherently split into the two spatial
modes $\left|a\right\rangle $ and $\left|b\right\rangle $ by means
of a polarization independent Mach-Zehnder interferometer, resulting
in the spatial state $\cos(\frac{\alpha}{2})\left|a\right\rangle +\sin(\frac{\alpha}{2})\left|b\right\rangle $.
The phase $\alpha$ is determined by the optical path-length difference
between the two interferometer arms. Next, the two spatial modes acquire
an additional phase difference $\phi$, resulting in the state\begin{equation}
\textrm{e}^{i\phi}\cos(\frac{\alpha}{2})\left|a\right\rangle +\sin(\frac{\alpha}{2})\left|b\right\rangle \label{eq:spatialState}\end{equation}

\noindent of the photonic qubit. In order to prepare a well-defined
state, precise control over the interferometric phases $(\alpha,\phi)$
is necessary. Therefore the optical path-length differences in the
interferometric setup are actively stabilized with the help of an
additional stabilization laser and an electronic feedback loop, allowing
measurement times of up to 24 hours. By inserting a rotatable glass
plate into the stabilization beam we can change these path-length
differences and thus precisely control the phase setting.

Next, in order to transfer the state given by Eq. (\ref{eq:spatialState})
onto the spin state of the atom, a Bell-state measurement in the polarization/spatial
mode Hilbert space of the photon is performed. This is done by combining
the two modes $\left|a\right\rangle $ and $\left|b\right\rangle $
on a polarizing beam-splitter and analyzing the photon polarization
in each output port (see Fig. \ref{cap:preparationPart}). The polarization
analyzer detects $\left|\pm45°\right\rangle =\frac{1}{\sqrt{2}}(\left|H\right\rangle \pm\left|V\right\rangle )$
polarized photons by means of four single photon counting Si avalanche
photo diodes (APD1..4). Since the PBS transmits horizontal $\left|H\right\rangle $
and reflects vertical $\left|V\right\rangle $ polarization, a coherent
superposition of orthogonal polarizations from both modes is necessary
to obtain $\left|\pm45°\right\rangle $ in the output of the PBS.
For example to get $\left|+45°\right\rangle $ in the PBS output with
detectors 1 and 2, $\left|H\right\rangle $ polarization has to be
transmitted from mode $\left|b\right\rangle $ and coherently added
to $\left|V\right\rangle $ polarization reflected from mode $\left|a\right\rangle $.
This corresponds to the Bell-state $\left|\Psi^{+}\right\rangle =\frac{1}{\sqrt{2}}(\left|V\right\rangle \left|a\right\rangle +\left|H\right\rangle \left|b\right\rangle )$.
Accordingly, the $\left|-45°\right\rangle $ polarization corresponds
to the $\left|\Psi^{-}\right\rangle =\frac{1}{\sqrt{2}}(\left|V\right\rangle \left|a\right\rangle -\left|H\right\rangle \left|b\right\rangle )$
state, while in the other output of the PBS the states $\left|\Phi^{\pm}\right\rangle =\frac{1}{\sqrt{2}}(\left|H\right\rangle \left|a\right\rangle \pm\left|V\right\rangle \left|b\right\rangle )$
are detected.

\noindent The Bell-state detection projects the atomic qubit onto
one of the four states

\begin{equation}
\begin{array}{c}
\left|\Phi_{1}\right\rangle =\textrm{e}^{i\phi}\cos(\frac{\alpha}{2})\left|\uparrow\right\rangle _{x}+\sin(\frac{\alpha}{2})\left|\downarrow\right\rangle _{x}\\
\left|\Phi_{2}\right\rangle =\textrm{e}^{i\phi}\cos(\frac{\alpha}{2})\left|\uparrow\right\rangle _{x}-\sin(\frac{\alpha}{2})\left|\downarrow\right\rangle _{x}\\
\left|\Phi_{3}\right\rangle =\textrm{e}^{i\phi}\cos(\frac{\alpha}{2})\left|\downarrow\right\rangle _{x}-\sin(\frac{\alpha}{2})\left|\uparrow\right\rangle _{x}\\
\left|\Phi_{4}\right\rangle =\textrm{e}^{i\phi}\cos(\frac{\alpha}{2})\left|\downarrow\right\rangle _{x}+\sin(\frac{\alpha}{2})\left|\uparrow\right\rangle _{x}\end{array}\label{eq: RSPStates}\end{equation}
where $\left|\uparrow\right\rangle _{x},\left|\downarrow\right\rangle _{x}=\frac{1}{\sqrt{2}}(\left|\uparrow\right\rangle _{z}\pm\left|\downarrow\right\rangle _{z})$.
State $\left|\Phi_{1}\right\rangle $ is already equivalent to the
photonic state from Eq. (\ref{eq:spatialState}). The states $\left|\Phi_{2}\right\rangle $,
$\left|\Phi_{3}\right\rangle $, and $\left|\Phi_{4}\right\rangle $
can be transformed into $\left|\Phi_{1}\right\rangle $ by applying
the operation $\hat{\sigma}_{x}$, $\hat{\sigma}_{y}$, or $\hat{\sigma}_{z}$,
respectively on the atom. 

After completion of the transfer of the state from the photon to the
atom we perform the analysis of the atomic state\cite{Volz06}. First,
a certain superposition of $\left|\uparrow\right\rangle _{z}$ and
$\left|\downarrow\right\rangle _{z}$ is transfered to a different
hyperfine level ($\left|F=2\right\rangle $) by means of a state-selective
STIRAP process. The polarization of the transfer pulse defines which
superposition is being transferred and thus allows the choice of the
measurement basis. The following hyperfine-state analysis measures
the fraction of population which was transferred by removing atoms
in the state $\left|F=2\right\rangle $ from the trap. This method
allows to analyze the state of the atom in any desired basis and thus
to reconstruct the density matrix of the state by combining measurements
in 3 complementary bases. The characterization of the entangled atom-photon
state with this method yields a fidelity of $87\%$.

\begin{figure*}[bt]
\begin{center}\includegraphics{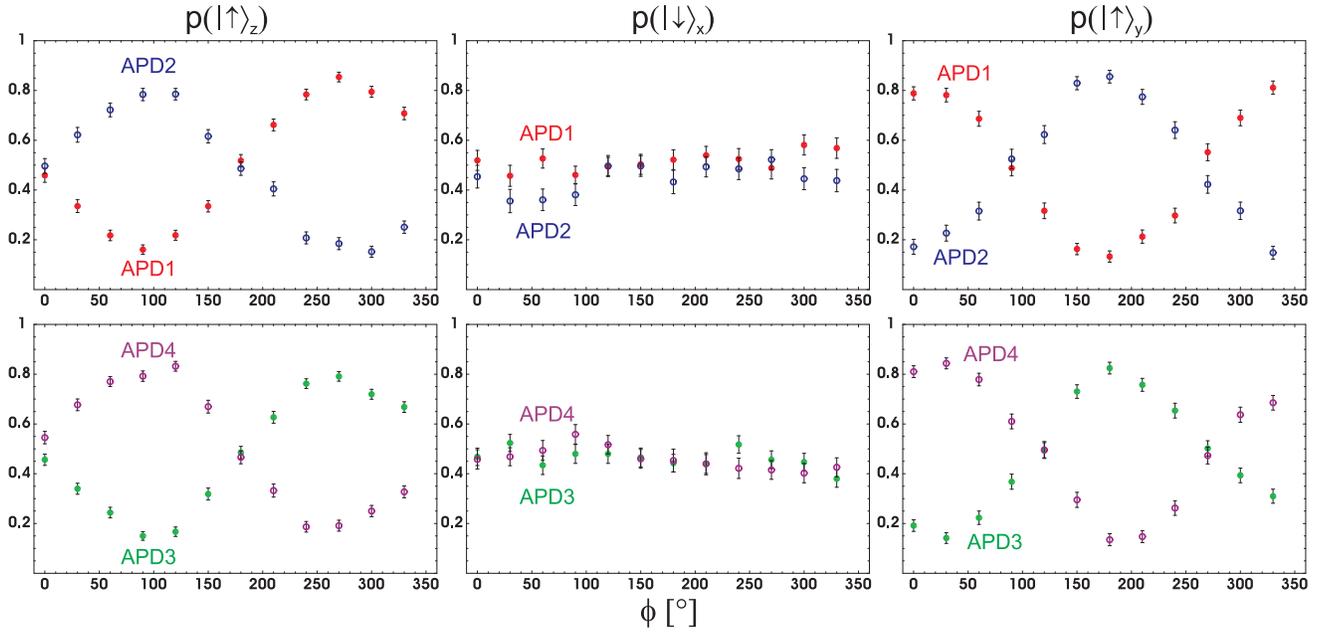}\end{center}

\caption{\label{cap:measurement01}Tomographic dataset of the prepared atomic
states for $\alpha=90°,\ \phi=0...330°$. The figures show the probability
p to find the atom in the state $\left|\uparrow\right\rangle _{z}$
(left), $\left|\downarrow\right\rangle _{x}=\frac{1}{\sqrt{2}}(\left|\uparrow\right\rangle _{z}-\left|\downarrow\right\rangle _{z})$
(middle) and $\left|\uparrow\right\rangle _{y}=\frac{1}{\sqrt{2}}(\left|\uparrow\right\rangle _{z}+i\left|\downarrow\right\rangle _{z})$
(right), respectively, after a photon detection in detector 1 (red,
filled) and 2 (blue, hollow) (upper row), 3 (green, filled) and 4
(magenta, hollow) (lower row). Each data-point is evaluated from 150-350
measurement processes from which we calculate the depicted statistical
errors (one standard deviation). The mean fidelity of the 12 states
prepared in this measurement is $82.6\%$. The acquisition of the
full dataset was realized within approximately three days at an event
rate of 10-20 per minute.}
\end{figure*}

\noindent In order to evaluate the performance of our preparation
scheme, we prepared different states of the atom by varying the phase
settings $(\alpha,\phi)$. Then we performed a full quantum state
tomography of the atomic qubit for each of the four detected Bell
states separately. Fig. \ref{cap:measurement01} exemplarily shows
a measurement where we set $\alpha=90°$ while rotating $\phi=0...330°$
in steps of $30°.$ Let us consider, e.g., the state which is prepared
when the photon is registered in detector APD1. This state can be
decomposed in three complementary bases as 

\begin{equation}
\begin{array}{c}
\left|\Phi_{1}\right\rangle =\cos(\frac{1}{2}(\phi+\frac{\pi}{2}))\left|\uparrow\right\rangle _{z}+i\cdot\sin(\frac{1}{2}(\phi+\frac{\pi}{2}))\left|\downarrow\right\rangle _{z}\\
=\frac{1}{\sqrt{2}}(\textrm{e}^{i\phi}\left|\uparrow\right\rangle _{x}+\left|\downarrow\right\rangle _{x})\\
=\cos(\frac{1}{2}\phi)\left|\uparrow\right\rangle _{y}+i\cdot\sin(\frac{1}{2}\phi)\left|\downarrow\right\rangle _{y}.\end{array}\end{equation}

\begin{figure}[bt]
\begin{center}\includegraphics{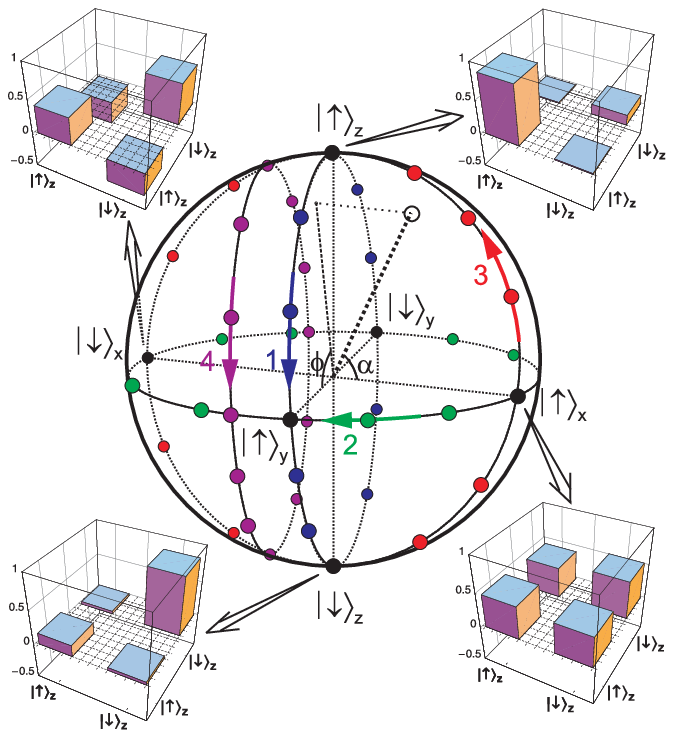}\end{center}

\caption{\label{cap:BlochSphere}Bloch-sphere representation of the states
prepared on the atomic qubit. The basis states in the equatorial plane
are defined as $\left|\uparrow\right\rangle _{x},\left|\downarrow\right\rangle _{x}:=\frac{1}{\sqrt{2}}(\left|\uparrow\right\rangle _{z}\pm\left|\downarrow\right\rangle _{z})$
and $\left|\uparrow\right\rangle _{y},\left|\downarrow\right\rangle _{y}:=\frac{1}{\sqrt{2}}(\left|\uparrow\right\rangle _{z}\pm i\left|\downarrow\right\rangle _{z})$.
The angles $(\alpha,\phi)$ can be interpreted as usual polar coordinates
with respect to the x-axis. The numbers 1-4 depict the corresponding
measurements from Table \ref{tab:Results}. The insets exemplarily
show measured density matrices of the atomic qubit (real part) for
four selected states.}
\end{figure}

\noindent While the projections of $\left|\Phi_{1}\right\rangle $
onto $\left|\uparrow\right\rangle _{x}$ and $\left|\downarrow\right\rangle _{x}$
are equal and constant, we observe a dependence on $\phi$ for the
projection onto $\left|\uparrow\right\rangle _{z},\left|\downarrow\right\rangle _{z}$
and $\left|\uparrow\right\rangle _{y},\left|\downarrow\right\rangle _{y}$.
By combining all three measurements we determined the density matrix
of each prepared atomic state. From this we derived the fidelity (which
is the probability to find the atom in the state expected from Eq.
(\ref{eq: RSPStates})) for each detector and every setting of $(\alpha,\phi)$.
The mean fidelity over all points and all four analyzed Bell-states
in this measurement is $82.6\%$. We performed 4 sets of measurements
of this kind preparing various states on different circles on the
Bloch sphere (see Fig. \ref{cap:BlochSphere}). Altogether, 42 different
states were prepared with a mean fidelity of $82.2\%$ (see Table
\ref{tab:Results}). 

\begin{table}[bt]
\begin{center}\begin{tabular}{|c|c|c|c|}
\hline 
\#&
$\alpha$&
$\phi$&
F\tabularnewline
\hline
\hline 
1&
$90°$&
$0..330°$&
$82.6\%\pm0.40\%$ \tabularnewline
\hline 
2&
$0..330°$&
$0°$&
$79.7\%\pm0.65\%$\tabularnewline
\hline 
3&
$0..330°$&
$90°$&
$84.2\%\pm0.45\%$\tabularnewline
\hline 
4&
$109.5°$&
$0..330°$&
$82.2\%\pm0.46\%$\tabularnewline
\hline
\end{tabular}\end{center}

\caption{\label{tab:Results}Summary of the experimental results. The table
shows the fidelity F, which is the probability of a successful state
transfer, averaged over all 4 detected Bell-states and all 12 points
within one measurement set.}
\end{table}

There are several sources of imperfections which affect the achieved
preparation fidelity. The most important factors are the limited purity
of the generated entangled atom-photon state and imperfections in
the atomic state detection, yielding together a reduced entanglement
fidelity of $87\%$. Taking into account this error source we get
a corrected fidelity of $\frac{0.82}{0.87}\approx94\%$ for the preparation/teleportation
process alone. This value is limited by the finite visibility of the
interferometer and Bell-state analyzer (about $96\%$), the mechanical
instability of the interferometer and the residual birefringence of
its components. The coherence of the prepared states decays on a time
scale of about $10\mathrm{\mu s}$ and does not influence the current
measurement. This decay is caused solely by dephasing due to magnetic
stray fields, resulting from instabilities of the magnetic field compensation.
Longer coherence times can be achieved by using an improved compensation
method.

The presented experiment demonstrates the faithful remote preparation
of arbitrary quantum states of a single atom without the need of a
direct interaction between the information carrier (photon) and the
quantum memory (atom). Our implementation uses a quantum teleportation
protocol to transfer the state of a photonic qubit onto the atom with
an average preparation fidelity as high as $82\%$. The long coherence
time of atomic ground states\cite{Cline94} makes such a system well
suited for future applications. In particular, the combination with
recent achievements in experiments with trapped atoms and ions makes
advanced schemes like quantum networks or the quantum repeater - almost
- state of the art. One could employ systems with a few atoms, where
some are used for tasks like computation, storage and entanglement
purification, others for establishing the communication link to neighboring
nodes via entanglement swapping. This way one profits from both, the
high fidelity and flexibility of quantum logic operations on atoms
or ions and the efficient transmission of photonic qubits that are
ideally suited for efficient long distance distribution of quantum
information.

This work was supported by the Deutsche Forschungsgemeinschaft and
the European Commission through the EU Project QAP (IST-3-015848)
and the Elite Network of Bavaria through the excellence program QCCC.

\end{document}